\documentclass[a4paper]{jpconf}
\usepackage{graphicx}
\usepackage{bm}
\usepackage{amsmath}
\usepackage{amssymb}

\newcommand{\be}{\begin{equation}}
\newcommand{\ee}{\end{equation}}
\newcommand{\bea}{\begin{eqnarray}}
\newcommand{\eea}{\end{eqnarray}}

\begin{document}
\title{Hairy Lovelock black holes and Stueckelberg mechanism for Weyl symmetry}

\author{Mariano Chernicoff $^1$, Gaston Giribet $^{2,3,4}$ and Julio Oliva$^{5}$}

\address{$^1$Departamento de F\'{\i}sica, Facultad de Ciencias, Universidad Nacional Aut\'onoma de M\'exico; A.P. 70-542, M\'exico D.F. 04510, M\'exico.

$^2$Departamento de F\'{\i}sica, Universidad de Buenos Aires and IFIBA - CONICET; Ciudad Universitaria, pabell\'on 1 (1428) Buenos Aires.

$^3$Universit\'e Libre de Bruxelles and International Solvay Institutes; Campus Plaine C.P. 231 B-1050, Bruxelles, Belgium.

$^4$Instituto de F\'{\i}sica, Pontificia Universidad Cat\'olica de Valpara\'{\i}so; Casilla 4950, Valpara\'{\i}so, Chile.

$^5$Departamento de F\'{\i}sica, Universidad de Concepci\'on; Casilla 160-C, Concepci\'on, Chile.}

\ead{mchernicoff@ciencias.unam.mx, gaston@df.uba.ar, julioolivazapata@gmail.com}

\begin{abstract}
Lovelock theory of gravity -and, in particular, Einstein theory- admits black hole solutions that can be equipped with a hair by conformally coupling the theory to a real scalar field. This is a secondary hair, meaning that it does not endow the black hole with new quantum numbers. It rather consists of a non-trivial scalar field profile of fixed intensity which turns out to be regular everywhere outside and on the horizon and, provided the cosmological constant is negative, behaves at large distance in a way compatible with the Anti-de Sitter (AdS) asymptotic. In this paper, we review the main features of these hairy black hole solutions, such as their geometrical and thermodynamical properties. The conformal coupling to matter in dimension $D>4$ in principle includes higher-curvature terms. These couplings are obtained from the Lovelock action through the Stueckelberg strategy. As a consequence, the resulting scalar-tensor theory exhibits a self-duality under field redefinition that resembles T-duality. Through this field redefinition, the matter content of the theory transforms into a Lovelock action for a dual geometry. Since the hairy black holes only exist for special relations between the dual Lovelock coupling constants, it is natural to compare those relations with the causality bounds coming from AdS/CFT. We observe that, while the lower causality bound is always obeyed, the upper causality bound is violated. The latter, however, is saturated in the large $D$ limit.  
\end{abstract}

\section{Introduction}

\vspace{0.2cm}

Lovelock theory \cite{Lanczos, Lovelock, Lovelock2} is the most general torsionless metric theory of gravity that yields field equations involving a rank-2 symmetric tensor of second order and covariantly conserved. Being the natural generalization of general relativity, Lovelock theory exhibits nice properties such as the absence of ghosts around physically sensible backgrounds \cite{Zwiebach, BoulwareDeser}; it propagates a single massless graviton despite of being a higher-curvature theory. The action of Lovelock theory consists of the Einstein-Hilbert action augmented with higher-curvature terms ${\mathcal O}(R^k)$ that involve specific contractions of the Riemann tensors with no derivatives of it. While for spacetime dimension $D\leq 4$ the theory reduces to Einstein theory, for $D>4$ it presents deviation from general relativity that typically emerge at distances comparable with the length scales defined by the dimensionful coupling constants of the higher-curvature terms. 

Lovelock theory is relevant in high-energy physics for different reasons: the higher-curvature terms in its action resemble those appearing in the next-to-leading order of string theory low energy effective action and in M-theory compactifications. More generally, it provides a tractable example to study the consistency of more general higher-curvature models \cite{Maldacena, Reall}. Being of second order, this theory is the perfect setup to investigate the effects of introducing higher-curvature corrections in different scenarios, and this is the reason why in the last years it has become the favorite model to study higher-curvature corrections within the context of AdS/CFT correspondence \cite{Myers, Edelstein, Paulos}; see \cite{Xian} and references therein. 

In reference \cite{Oliva1}, it has been shown that Lovelock theory admits a natural generalization that leads to write down the most general conformally invariant single scalar field theory with second order field equations in $D$ dimensions. This is in the same spirit as Horndeski and Galileon theories \cite{Horndeski, Galileon}. In reference \cite{Giribet1}, hairy black hole solutions to Lovelock gravity coupled to such conformally invariant matter theory were found for special curves of the space of coupling constants. The scalar hair of these black holes is a {\it secondary hair}; that is, the scalar field does not endow the black hole with new quantum numbers, but its intensity turns out to be fixed in terms of some of the coupling constants in the action. The scalar field configuration is regular everywhere outside and on the horizon, and behaves at large distance in a way that respects the AdS asymptotic -here we will consider the case of negative cosmological constant-. In this paper, we will study different aspects of these hairy black hole solutions. We will review their main geometrical and thermodynamical properties. We will also study a duality that Lovelock theory conformally coupled to the scalar field exhibits under field redefinitions. Through such a field redefinition, the matter content of the theory transforms into a Lovelock action for a {\it dual} geometry. Since  the hairy black holes exist only for special relations among the dual Lovelock coupling constants, we will compare those relations with the causality bounds coming from AdS/CFT. We will show that, while the lower causality bound is always obeyed, the upper causality bound is violated. Nevertheless, the latter is saturated in the large $D$ limit.

\section{Higher-curvature gravity and conformal couplings}
\label{theory}
\vspace{0.2cm}
\subsection{Lovelock theory}
\vspace{0.2cm}

Lovelock theory is defined by the action
\begin{equation}\label{aL}
I_{\{a\}}[g]=\int_{\mathcal M} d^Dx \sqrt{-g}\sum^{[\frac{D-1}{2}]}_{k=0} a_k  \frac{(2k)!}{2^k}
\delta^{\mu_1}_{[\alpha_1}
\delta^{\nu_1}_{\beta_1} \ldots\ \delta^{\mu_k}_{\alpha_k} \delta^{\nu_k}_{\beta_k ]}
\  R^{\ \ \ \ \alpha_1\beta_1}_{\mu_1\nu_1}\dots R^{\ \ \ \ \alpha_k\beta_k}_{\mu_k\nu_k}
  \ ,
\end{equation}
where $R^{\ \mu}_{\nu\ \lambda\delta}$ is the Riemann curvature tensor, the symbol $[n]$ in the sum stands for the integer part of $n$, other brackets refer to antisymmetrization. The collection of real numbers $a_0, a_1, \ldots  a_k$ in (\ref{aL}), to which the subindex $\{a\}$ on the left hand side of (\ref{aL}) refers, are coupling constants of the theory that are in principle independent. However, being dimensionful coupling constants, it is natural to think of them as being given by a unique fundamental length scale $\ell_P$ with $a_k\sim \ell_P^{2k-D}$. In (\ref{aL}), we are omitting boundary terms, cf. \cite{Galante1} (see also \cite{Mann}). This action yields the most general theory of gravity whose fields equations are of second order in the metric. These field equations read 
\begin{align} \label{elG}
\mathcal{G}_{\mu}^{\nu}   \equiv \sum_{k=0}^{\left[  \frac{D-1}{2}\right]  } a_k \frac{(2k)!}{2^{k+1}}
\delta^{\nu }_{[\mu }
\delta^{\alpha_1}_{\beta_1} \ldots\ \delta^{\alpha_{2k}}_{\beta_{2k} ]}
R_{\ \ \ \ \alpha_{1}\alpha_{2}}^{\beta_{1}\beta_{2}}\ldots R_{\ \ \ \ \ \ \ \alpha
_{2k-1}\alpha_{2k}}^{\beta_{2k-1}\beta_{2k}}\ =0 \,,
\end{align}
which, indeed, are of second order in the metric, although not of degree one in its second derivatives. Therefore, (\ref{elG}) can be regarded as the natural generalization of Einstein equations to higher dimensions. 


\subsection{Conformal couplings}
\vspace{0.2cm}

Among the different ways to couple matter to Lovelock theory, there is one that results particularly interesting because it provides a tractable model that exhibits conformal symmetry. In dimension $D\geq 5$ a conformal invariant action for a scalar field may in principle include higher-curvature couplings. To write down the most general action of this sort, one may resort to the Stueckelberg strategy: Lovelock theory is not Weyl invariant. However, one can define the rescaled metric $\tilde{g}_{\mu \nu } = \varphi^{-2/s}g_{\mu \nu}$, with $s$ being a real constant different from zero and $\varphi $ being a real scalar field. Then, one considers the Riemann curvature tensor associated to $\tilde{g}_{\mu \nu }$,
\begin{equation}\label{Stensor}
S_{\mu\nu} ^{\ \ \lambda\delta}\equiv \varphi^2R^{\ \ \lambda\delta}_{\mu\nu}+\frac{4}{s}\varphi\delta^{[\lambda}_{[\mu}\nabla_{\nu]}\nabla^{\delta]}\varphi +\frac{4(1-s)}{s^2}\delta^{[\lambda}_{[\mu}\nabla_{\nu]}\varphi\nabla^{\delta]}\varphi -\frac{2}{s^2}\delta^{[\lambda}_{[\mu}\delta^{\delta]}_{\nu]}\nabla_{\rho}\varphi\nabla^{\rho}\varphi ,
\end{equation}
which happens to transform homogeneously under the Weyl transformation
\begin{equation}\label{conformaltrans}
g_{\mu \nu }\rightarrow e^{2\sigma }g_{\mu \nu }, \qquad \varphi \rightarrow e^{s\sigma }\varphi \,
\end{equation}
with $\sigma $ being a function of the spacetime coordinates. More precisely, under the rescaling (\ref{conformaltrans}) tensor (\ref{Stensor}) transforms as 
\begin{equation}
S_{\mu\nu} ^{\ \ \lambda\delta} \rightarrow e^{2(s-1)\sigma }S_{\mu\nu} ^{\ \ \lambda\delta} \,.
\end{equation}

This permits to construct a conformal invariant action for the compensator field $\varphi $ by simply replacing $R^{\ \mu}_{\nu \ \ \lambda\delta}$ by $S^{\ \mu}_{\nu \ \ \lambda\delta}$ in (\ref{aL}). Then, the general action of Lovelock theory coupled to conformal matter takes the form \cite{Oliva1}
\begin{align}\label{theoryaction}
I[g,\varphi ]=\int_{\mathcal M} d^Dx \sqrt{-g}\sum^{[\frac{D-1}{2}]}_{k=0}\frac{(2k)!}{2^k}
\delta^{\mu_1}_{[\alpha_1}
\delta^{\nu_1}_{\beta_1} \ldots\ \delta^{\mu_k}_{\alpha_k} \delta^{\nu_k}_{\beta_k ]}
\Big{(} & a_k \ R^{\ \ \alpha_1\beta_1}_{\mu_1\nu_1}\dots R^{\ \ \alpha_k\beta_k}_{\mu_k\nu_k}
+\nonumber\\
& b_k\varphi^{m_k}  S^{\ \ \alpha_1\beta_1}_{\mu_1\nu_1}\dots S^{\ \ \alpha_k\beta_k}_{\mu_k\nu_k}  \Big{)} \,,
\end{align}
where $a_k$ and $b_k$ represent coupling constants that are in principle arbitrary -only constrained by consistency conditions such as unitarity and causality-, and where $
m_k=(2k-D)/s-2k $. The parameter $s$ corresponds to the conformal weight of the scalar field. The choice $s=1-D/2$ leads to the standard action of a conformally coupled scalar field theory
\begin{equation}
I[g,\varphi ]=\int_{\mathcal M} d^Dx \sqrt{-g} \Big( \frac{1}{16\pi G} R - \frac{\Lambda}{8\pi G} - (\nabla \varphi )^2 - \frac{(D-2)}{4(D-2)}R\varphi^2 - \lambda \varphi^{2D/(D-2)}+ ... \Big) \nonumber
\end{equation}
where we have chosen $b_1=-1$ and denoted $a_0=\Lambda/(8\pi G)$, $a_1=1/(16\pi G)$, $b_0=-\lambda $ as usual. The $D$-dimensional Planck length is given by $\ell_P= G^{1/(D-2)}$. The ellipses stand for terms with more than two derivatives on the metric and on $\varphi $. The latter, however, do not yield higher-order field equations. In fact, the new conformal couplings involve higher-curvature terms together with higher-derivatives of the scalar field, all arranged in a way that the field equations turn out to be conformally invariant as well as of second order \cite{Oliva1}. The field equations coming from varying (\ref{theoryaction}) with respect to the metric can be written in the form
\begin{equation}\label{E210}
\mathcal{G}_{\mu\nu}=8\pi \ell^{D-2}_P T_{\mu\nu}  \,,
\end{equation}
with $\mathcal{G}_{\mu\nu}$ given by (\ref{elG}) and $T_{\mu\nu}$ given by
\begin{align}
T_{\mu}^{\nu}   =\sum_{k=0}^{\left[  \frac{D-1}{2}\right]  }\frac{(2k)!a_1b_{k}%
}{2^{k}}\varphi^{m_k}
\delta^{\nu }_{[\mu}
\delta^{\alpha_1}_{\beta_1} \ldots\ \delta^{\alpha_{2k}}_{\beta_{2k} ]}
S_{\ \ \ \ \alpha_{1}\alpha_{2}}^{\beta_{1}\beta_{2}%
} \ldots S_{\ \ \ \ \ \ \ \ \alpha_{2k-1}\alpha_{2k}}^{\beta_{2k-1}\beta_{2k}} \,.
\end{align}

From this, we easily verify that, despite the presence of higher-curvature terms in the action, equations (\ref{E210}) are of second order in the fields. The same happens with the equation for the scalar field $\varphi $, which reads
\begin{equation}\label{onyel}
\sum_{k=0}^{[\frac{D-1}{2}]} \frac{ \left(2k-D\right) b_{k} }{ 2^{k}s } \varphi^{m_{k}-1} 
\delta^{\mu_1}_{[\alpha_1 }
\delta^{\nu_1}_{\beta_1} \ldots\ \delta^{\mu_{k}}_{\alpha_{k} } \delta^{\nu_{k}}_{\beta_{k} ]}
S^{\ \ \ \ \alpha_1\beta_1}_{\mu_1\nu_1} \dots S^{\ \ \ \ \alpha_k\beta_k}_{\mu_k\nu_k}=0 \,.
\end{equation}

It is possible to show that the trace of $T_{\mu\nu}$ vanishes on-shell. This is consistent with the conformal invariance of the matter action.

\section{Hairy black hole solution}
\label{bhsolution}
\vspace{0.2cm}

In reference \cite{Giribet1}, explicit black hole solutions to field equations (\ref{E210}) have been found for arbitrary dimension $D>4$, for horizons of either positive or negative curvature, and for arbitrary sign of the cosmological constant. Here, we will focus our attention to asymptotically AdS solutions with horizons of spherical topology. 

There are special relations between the coupling constants $b_0, b_1, ...\ b_{[(D-1)/2]}$ for which the static spherically symmetric solution to the field equations (\ref{E210})-(\ref{onyel}) is compatible with the ansatz
\begin{equation}\label{solutionhigher}
ds^2=-F(r)dt^2+\frac{ 1}{F(r)}dr^2+r^2d\Omega^2_{D-2}  ,
\end{equation}
where $r \in \mathbb{R}_{> 0}$, $t\in \mathbb{R}$, and where $d\Omega^2_{D-2}$ is the metric of a $(D-2)$-dimensional Euclidean space of constant curvature. Here, we will consider this space to be the unit $(D-2)$-sphere. In that case, the metric function $F(r)$ is root of the polynomial equation \cite{Wheeler}
\begin{equation}\label{polynomialeq}
\sum^{[\frac{D-1}{2}]}_{k=0}a_k\frac{(D-1)!}{(D-2k-1)!}\Big{(}\frac{1-F(r)}{r^2}\Big{)}^k=\frac{M(D-1)}{\text{Vol}_{\Omega}r^{D-1}}+\frac{Q_0}{r^D} \,,
\end{equation}
where $\text{Vol}_{\Omega}={2\pi^{(D-1)/2}}/{\Gamma ((D-1)/2)}$ is the volume of the unit $(D-2)$-sphere.

Choosing $s=-1$, the solution for the scalar field takes the simple form  
\begin{equation}\label{solutionhigherscalar}
\qquad \varphi(r)=\frac{N}{r} ,
\end{equation}
with $N$ satisfying the following constraints
\begin{equation}\label{const1}
\sum^{[\frac{D-1}{2}]}_{k=1} k\tilde{b}_k N^{2-2k}=0 \ , \ \ \ \sum^{[\frac{D-1}{2}]}_{k=0} (D(D-1)+4k^2)\tilde{b}_k N^{-2k}=0 ,
\end{equation}
with $\tilde{b}_k= {b_k(D-1)!}/{(D-2k-1)!} $. These constraints also give a specific relation between the different coupling constants $b_0, b_1, ...\ b_{[(D-1)/2]}$ for metric (\ref{solutionhigher})-(\ref{polynomialeq}) to be a solution.

$M$ in (\ref{polynomialeq}) is an arbitrary integration constant that corresponds to the black hole mass. In contrast, the constant $Q_0$ that weights the backreaction of the scalar field is fixed in terms of the coupling constants through the equation
\begin{equation}\label{charge}
Q_0=\sum^{[\frac{D-1}{2}]}_{k=0} (D-2k-1)\tilde{b}_k N^{D-2k} \,.
\end{equation}
This means that the conformal field $\varphi $ corresponds to a secondary hair. The scalar field configuration remains finite everywhere outside and on the horizon, and it decays at large distance with a power law that depends on $s$ (we considered in (\ref{solutionhigherscalar}) the value $s=-1$). The particular dependence on $Q_0$ in the metric can be understood as the backreaction on the geometry produced by the scalar field energy density 
\begin{equation}\label{jhgflkhf}
T^{\ 0}_0 (r) = -\frac{Q_0}{(D-1) r^D} ,
\end{equation} 
which is positive provided $Q_0<0$. Equation (\ref{polynomialeq}) receives additional contributions on the right hand side when the black hole is charged under $U(1)$ gauge fields; see (\ref{Hoy23}) below. 

For illustrative purposes it is enough to consider only up to ${\mathcal O}(R^2)$ terms in the purely gravitational part of the action; therefore, hereafter we set $a_{k>2}=0$. In such case, equation (\ref{polynomialeq}) notably simplifies and yields
\begin{equation}
F(r) = 1+\frac{r^2}{4\tilde{\alpha}}  \left( 1 - H(r)\right) \,, \label{LaF}
\end{equation}
with
\begin{eqnarray}
H^2(r) &=& 1-\frac{8\tilde{\alpha}a_0}{(D-1)(D-2)a_1}+\frac{8\tilde{\alpha}M}{\text{Vol}_{\Omega}(D-2)a_1r^{D-1}}+ \nonumber \\ 
&&\frac{8\tilde{\alpha}Q_0}{(D-1)(D-2)a_1r^D}  
- \frac{8\tilde{\alpha } Q_1^2}{(D-2)(D-3)a_1 \text{Vol}_{\Omega }r^{2D-4}} 
 \,, \nonumber
\end{eqnarray}
where we have defined $\tilde{\alpha} =a_2(D-4)(D-3)/(2a_1)$. This is a generalization of the Boulware-Deser solution \cite{BoulwareDeser} and of its electrically charged generalization \cite{Wiltshire}; see also \cite{Garraffo}. $Q_1$ is the black hole charge under a minimally coupled Abelian gauge field with electric solution 
\begin{equation}\label{Hoy23}
A(r) = \frac{Q_1 }{(D-3)r^{D-3}} dt. 
\end{equation}

The equation for $H^2(r)$ given above is quadratic, allowing in principle for the two signs of $H(r)$. This accounts for the two roots of $F(r)$ in the quadratic version of equation (\ref{polynomialeq}). Nevertheless, only the positive root corresponds to a physically sensible solution \cite{BoulwareDeser}, yielding the following behavior for small $\tilde{\alpha }$
\begin{eqnarray}
F(r)&\simeq & 1-\frac{16\pi G M}{\text{Vol}_{\Omega } (D-2) r^{D-3}}+\frac{16\pi G Q_1^2}{\text{Vol}_{\Omega } (D-2)(D-3) r^{2D-6}}-
\nonumber \\
&&\frac{16\pi G Q_0}{(D-1)(D-2)r^{D-2}}- \frac{2\Lambda r^2}{(D-1)(D-2)} + {\mathcal O}(\tilde{\alpha }) \,, \nonumber
\end{eqnarray}
where again we have denoted $a_0 = -{\Lambda }/({8\pi G})$ and $a_1=1/(16\pi G)$. At first order in $\tilde{\alpha}$ this corresponds to the general relativity (GR) solution, with the second term on the right hand side being the Newtonian potential of the Schwarzschild-Tangherlini solution, the third and four being the Reissner-Nordstr\"{o}m and the CFT matter contributions, respectively. The fifth term corresponds to the cosmological constant. 


From (\ref{LaF}), one can write the mass as a function of the horizon radius $r_+$, namely
\begin{equation}
M= \frac{\text{Vol}_{\Omega } }{16\pi G}
\Big{[}
- \frac{2\Lambda  r_+^{D-1}}{(D-1)}
+ (D-2)r_+^{D-3}
+ 2(D-2)\tilde{\alpha }r_+^{D-5}+
\frac{16\pi G Q_1^2}{(D-3)\text{Vol}_{\Omega }r_+^{D-3}} - \frac{16\pi GQ_0}{(D-1)r_+} \Big{]}  \label{HJK}
\end{equation}
which, if $Q_1=0$ and $Q_0>0$, is not bounded from below due to the presence of the last term. This pathology compels to consider only black holes with ${Q}_0\leq 0$. It is worth mentioning that, even if the $U(1)$ charge is present and screens the ultraviolet divergence due to ${Q}_0$, the solution with ${Q}_0>0$ still presents other pathologies, such as negative entropy configurations \cite{Galante1}. Therefore, hereafter we will restrict ourselves to the case ${Q}_0<0$.

\section{Black hole thermodynamics}
\vspace{0.2cm}

Now, let us discuss the thermodynamical properties of these hairy black holes. We will focus our attention on the effects produced by the presence of the contribution $Q_0\neq 0$. Hereafter we consider $Q_1 =0$. In this case, the Hawking temperature of the black hole solution (\ref{LaF}) is given by
\begin{eqnarray}
T = \frac{1}{4\pi(4\tilde{\alpha}+r^2_+)} \Big{[}
-\frac{2\Lambda r_+^{3}}{(D-2)}
+(D-3)r_+
+\frac{2(D-5)\tilde{\alpha }}{r_+ } +\frac{16\pi GQ_0}{(D-1)(D-2)r_+^{D-3}}  
\Big{]} \label{temperature}
\end{eqnarray}

The temperature presents different features depending on the parameters and it is necessary to consider different cases. At large $r_+$, the temperature correctly reproduces the behaviour of black holes in GR. The difference appears for small values of $r_+$. At short distances, provided $Q_0\neq 0 $, the leading term of (\ref{temperature}) goes like like $T\sim  Q_0/{(4 \tilde{\alpha}r^{D-3}_+ +r^{D-1}_+)}$. However, it is important to take into account that physically sensible solutions correspond to $Q_0 <0$. Therefore, there is a minimum value of $r_+$ for which $T\geq 0$. This is related to the fact that hairy black holes with sufficiently small radius exhibit positive specific heat. This leads to a slowed evaporation that yields remnants. This endpoint corresponds to extremal solutions with zero temperature. 

From the Euclidean action one can compute the entropy
\begin{equation}\label{entropy}
S =  4\pi \, \text{Vol}_{\Omega}(D-2) \, a_1 \, \left( \frac{r^{D-2}_+}{D-2}+\frac{4\tilde{\alpha}r^{D-4}_+}{D-4} \right) +S_0 = \frac{A}{4G}+{\mathcal O}(\tilde{\alpha } r^{D-4}) ,
\end{equation}
which obeys the first law of black hole thermodynamics with the mass and the temperature computed above. While the first term in (\ref{entropy}) corresponds to the Bekenstein-Hawking area law, the entropy formula also have other contributions. On the one hand, there are corrections due to the Lovelock terms in the pure gravity part of the action. 
Notice also that there is a constant contribution $S_0$ which depends on the couplings $b_0, b_1, ...\ b_{[(D-1)/2]}$. Such a constant contribution, puzzling from the physical point of view, is a recurrent feature of theories that contain higher-curvature terms. In the model discussed here, the value of the constant $S_0$ depends on the value of $Q_0$, and also its sign does. As a consequence, in order to avoid negative values of entropy for positive $r_+$, it is necessary to restrict the analysis to the cases $Q_0 \leq 0$, $\tilde{\alpha } \geq 0$. The presence of $S_0$ in the entropy formula has implications for thermodynamical (in)stability of the black holes as well as for other features such as their inner black hole properties.

\section{Dual geometry and causality bounds}
\vspace{0.2cm}

Because the matter part of the action has been constructed by means of a Stueckelberg mechanism from the Lovelock action, the former can also be thought of as a theory of pure gravity for the dual metric $\tilde{g}_{\mu\nu }$. In other words, the action of the full theory (\ref{theoryaction}) can be written as \cite{chernicoff}
\begin{equation}
I[g ,\varphi ] = I_{\{a\} }[g] + I_{\{b\} }[\varphi^{-2/s}g] , 
\end{equation}
or equivalently as
\begin{equation}
I[g ,\varphi ] = I_{\{b\} }[\tilde{g}] + I_{\{a\} }[{\varphi}^{+2/s}\tilde{g}]  . \label{duolingo} 
\end{equation}

This means that for special choices of the couplings constants $a_k$ and $b_k$ the theory turns out to be self-dual under the transformation
\begin{equation}
g_{\mu\nu } \leftrightarrow \tilde{g}_{\mu\nu}= \varphi^{-2/s} g_{\mu \nu } \ , \ \ \ \ \ \varphi \leftrightarrow \tilde{\varphi } = {\varphi }^{-1} \ .
\end{equation}

Constraints (\ref{const1}) admit particular solutions with $b_{k}=0$ for all $k>2$. This solution is
\begin{equation}
b_0 = -b_1\frac{N^2}{2(D-3)(D-4)} , \ \ \ b_2 = -b_1\frac{(D-2)(D^2-D-8)}{2D N^2} \ , \label{solucione}
\end{equation}
which yields
\begin{equation}
\frac{b_0 b_2}{b_1^2} = \frac{(D-2)(D^2-D-8)}{4D(D-3)(D-4)} \ . \label{Causalon1}
\end{equation}

These couplings $b_k$ can be thought of as the Lovelock couplings $\tilde{a}_k$ of the dual quadratic gravity theory (\ref{duolingo}) for the metric $\tilde{g}_{\mu \nu }$. The dual theory admits asymptotically (A)dS solutions with curvature radius of order $|b_1/b_0|^{1/2}$. Therefore, it is natural to compare (\ref{Causalon1}) with the causality constraint coming from AdS/CFT in $D$ dimensions. For the dual CFT not to propagate perturbations with superluminal velocities, the couplings of the quadratic Lovelock action must lie within the window \cite{Edelstein, Paulos, GeSin}
\begin{equation}
-\frac{(D-3)(3D-1)}{4(D+1)^2} \leq \frac{\tilde{a}_0 \tilde{a}_2}{\tilde{a}_1^2} \leq \frac{(D-3)(D-4)(D(D-3)+8)}{4(D^2-5D+10)^2} . \label{causalon}
\end{equation}

Comparing (\ref{causalon}) with (\ref{Causalon1}) in arbitrary dimension $D\geq 5$ one finds that the black hole solution exists for choices of the coupling constants $b_k$ such that the dual quadratic gravity theory only obeys the lower bound in (\ref{causalon}) but does not obey the upper bound. The upper bound, however, is met in the large $D$ limit, when ${b_0 b_2}/{b_1^2}\sim 1/4 + {\mathcal O}(1/D^2)$. In order to not to misinterpret this observation, it is worth pointing out that the violation of the upper causality constraint (\ref{causalon}) does not imply a priori the instability of the hairy black hole solution. This is because the dual geometry $\tilde{g}_{\mu \nu}$ that corresponds to the original solution (\ref{solutionhigher}) is not the AdS$_D$ vacuum that the dual theory has, but a different solution that asymptotes to $\mathbb{R}^{1,1} \times S^{D-2}$ with the radius of the $(D-2)$-sphere given by $N$. It is also interesting that the hairy black hole solution with $\Lambda =0$ corresponds to a dual geometry that asymptotes to AdS$_2\times S^{D-2}$.

It may be also interesting to compare the expressions (\ref{Causalon1}) and (\ref{causalon}) with other special value for the combination ${b_0 b_2}/{b_1^2}$; namely,
\begin{equation}
\frac{b_0 b_2}{b_1^2} = \frac{(D-2)(D-1)}{4(D-4)(D-3)} \label{Proco34}
\end{equation}    
which corresponds to a curve on the space of the coupling constants for which the dual quadratic Lovelock theory has a single (A)dS vacuum and not two different effective cosmological constants. In $D=5$ dimensions this coincides with the Chern-Simons theory (the special point $\lambda = 1/4$ in the notation employed in the AdS/CFT analysis, e.g. \cite{Myers, Edelstein, Paulos}), which exhibits peculiar properties. In the large $D$ limit of (\ref{Proco34}) one finds again the behaviour ${b_0 b_2}/{b_1^2}\sim 1/4 + {\mathcal O}(1/D)$. It would be interesting to further explore the properties that the solutions of this higher-curvature scalar-tensor theory exhibit in the large $D$ limit \cite{Emparan}.

\ack
\vspace{0.2cm}
The authors thank Jos\'e Edelstein, Mario Galante, Andr\'es Goya, Matias Leoni, Guillem P\'erez-Nadal, and Sourya Ray for discussions. This work has been supported by FONDECYT 1141073. and it has also been partially funded by FNRS-Belgium (convention FRFC PDR T.1025.14 and convention IISN 4.4503.15), by the Communaut\'{e} Fran{c}aise de Belgique through the ARC program and by a donation from the Solvay family. MC is partially supported by Mexico's National Council of Science and Technology (CONACyT) grant 238734 and DGAPA-UNAM grant IN113115. GG is partially supported by CONICET of Argentina through grant PIP0595-13.

\end{document}